\documentclass[english,11pt,aps,prd,a4paper,preprintnumbers,floatfix,nofootinbib,showpacs,superscriptaddress, notitlepage]{revtex4-1} 

\newcommand{\AddrUNAM}{Instituto de F\'isica, Universidad Nacional Aut\'onoma de M\'exico, A.P. 20-364, Ciudad de M\'exico 01000, M\'exico.}

\usepackage{color}
\usepackage{graphicx}
\usepackage{caption}
\usepackage{subcaption}
\usepackage{amsmath}
\usepackage{amssymb}
\usepackage[colorlinks=true,citecolor=darkred,urlcolor=darkred, pdfborder={0 0 0}]{hyperref}
\usepackage[normalem]{ulem}
\definecolor{darkred}{rgb}{0.6,0,0}
\definecolor{drkgrn}{RGB}{0, 51, 0}
\usepackage{soul}

\definecolor{linkcolor}{rgb}{0,0,0.5}

\def\gsim{\raise0.3ex\hbox{$\;>$\kern-0.75em\raise-1.1ex\hbox{$\sim\;$}}}
\def\lsim{\raise0.3ex\hbox{$\;<$\kern-0.75em\raise-1.1ex\hbox{$\sim\;$}}}

\def\beqn#1{\begin{equation}\label{#1}}
\def\eeqn{\end{equation}}

\def\beqa#1{\begin{eqnarray}\label{#1}}
\def\eeqa{\end{eqnarray}}

\def\Z2{$\mathcal{Z_2}$}

\newcommand {\ignore}[1]{}

\begin{document}

\title{\boldmath Dirac neutrinos from Peccei-Quinn symmetry: two examples}
\author{Leon M.G. de la Vega}\email{leonm@estudiantes.fisica.unam.mx}\affiliation{\AddrUNAM}
\author{Newton Nath}\email{newton@fisica.unam.mx}\affiliation{\AddrUNAM}
\author{Eduardo Peinado} \email{epeinado@fisica.unam.mx}\affiliation{\AddrUNAM}

\vspace{2.0cm}
\begin{abstract}
{\noindent
We aim to explain the nature of neutrinos using Peccei-Quinn symmetry. We discuss two simple scenarios,  one based on a type-II Dirac seesaw  and the other in a one-loop neutrino mass generation, which solve the strong CP problem and naturally lead to Dirac neutrinos. In the first setup latest neutrino mass limit gives rise to axion which is in the reach of conventional searches. Moreover, we have both axion as well as WIMP dark mater for our second set up.    
}
\end{abstract}

\maketitle


\section{Introduction}

Over the past few decades there has been remarkable progress in the field of particle physics. The discovery of neutrino oscillations \cite{Tanabashi:2018oca} provides a major milestone to
understand some intriguing aspects of neutrinos, which clarify the fact that neutrinos possess non-zero mass and their different flavors are mixed.
Apart from these, various observed phenomena provide some hints for the  existence of a non-baryonic form of matter, known as dark matter \cite{Bertone:2004pz}, in the Universe.  
Both of these issues are the most serious drawbacks of the Standard Model (SM) of particle physics.
Thus, they provide a clear evidence of new physics beyond the SM. 
Besides these, the SM also fails on  shedding  light on the strong CP problem of QCD, suggested by an experimental bound of the electric dipole moment of a neutron \cite{Kim:1986ax}. Peccei-Quinn (PQ) \cite{Peccei:1977hh} symmetry has been the most appreciated approach to explain the strong CP problem. 
The  PQ symmetry predicts the existence of the associated pseudo-Nambu-Goldstone (pNG) boson, the axion \cite{Wilczek:1977pj,Weinberg:1977ma} which can be a good cold dark matter candidate  \cite{Dine:1982ah,Abbott:1982af,Preskill:1982cy,Davis:1986xc}. 
Another puzzling challenge in the neutrino sector is whether neutrinos are Dirac or Majorana particles.  Despite the ongoing experimental effort on the search for the neutrinoless double beta decay~\cite{KamLAND-Zen:2016pfg}, which if observed will indicate the Majorana nature of neutrinos~\cite{Schechter:1981bd}, no signal of this process has been detected.  
Connecting these seemingly unrelated puzzles with the smallness of the neutrino masses is the scope of the present manuscript.

Axion models are mainly categorized into two classes, depending on whether quarks are charged under PQ symmetry or not, namely, DFSZ \cite{Dine:1981rt,Zhitnitsky:1980tq} and KSVZ \cite{Kim:1979if,Shifman:1979if}.
In axion models, where  quarks carry PQ charge, one needs two Higgs doublets $H_u$ and $H_d$ both charged under PQ symmetry in such a way that they couple to the PQ field $\sigma$ (singlet under the SM gauge group). There are two possibilities for such a coupling, namely it can be trilinear or quartic.  When the coupling is quartic, the spontaneous breaking of the Peccei-Quinn symmetry can be connected to the breaking of lepton number by two units~\cite{Mohapatra:1982tc,Langacker:1986rj} leading to Majorana neutrinos \cite{Schechter:1980gr}.

Recently, it has been shown that in a specific DFSZ axion scenario that  neutrinos can only be Dirac particles~\cite{Peinado:2019mrn}.  In order to explain the small (effective) neutrino Yukawa coupling, the tree-level coupling with the Higgs is forbidden by means of the PQ symmetry and allowed at the dimension-5 level. The PQ field plays a role in generating the Dirac neutrino mass which is proportional to the PQ breaking scale. In ~\cite{Peinado:2019mrn}  the Dirac neutrino masses were generated through a type-I Dirac seesaw. Scenarios where the neutrino mass mechanism and the PQ symmetry breaking are related has  attracted attention\footnote{For alternative solution to the strong CP problem connected with neutrino masses, see for instance~\cite{Hung:2017juv,Hung:2017exy,Carena:2019nnd}.}, both in the Majorana \cite{Dias:2005dn,Dasgupta:2013cwa,Bertolini:2014aia,Ahn:2015pia,Suematsu:2017kcu,Ma:2017zyb,Reig:2018yfd} and Dirac \cite{Gu:2016hxh,Baek:2019wdn,Chen:2012baa,Carvajal:2018ohk} frameworks.

In this work, we continue with the same approach by giving two different ways for naturally generating small Dirac neutrino masses.  In the first case, we extend the idea by using a type-II seesaw with exactly the same PQ charges for the RH neutrino and by including an extra Higgs doublet to give the small Dirac neutrino masses. 
 An explanation of the small Dirac neutrino masses within the DFSZ  axion model has been pointed in \cite{Gu:2016hxh} for type - I, II, and III seesaw framework. It is worthwhile to point out here that 
they have considered quartic coupling which is responsible to generate the Dirac neutrino masses, whereas  our focus is on trilinear coupling to explain the same within the type - II seesaw model. Moreover,  quartic coupling between Higges and axion has also been considered by the authors, while we concentrate on the trilinear coupling, which is crucial to demonstrate the Diracness of neutrinos.
 Recently, Ref.~\cite{Baek:2019wdn} has also explained the Dirac neutrino masses considering both the DFSZ and KSVZ class of models within the type-II Dirac seesaw formalism. However, the major difference lies in the choices of PQ-charges of Refs.~\cite{Gu:2016hxh,Baek:2019wdn}, which help us to show that even at the $n^{th}$ order one can not construct $U(1)_{PQ}$ invariant Majorana neutrino mass terms, hence neutrinos must be Dirac in nature. 
Another option we explore  is to generate neutrino masses at one loop level \footnote{Following the reasoning in~\cite{Peinado:2019mrn} it can also be extended to more than one loop but this is not the scope of the present manuscript.}. 
 In the second case, we kept exactly the same PQ charges for the right-handed (RH) neutrino as well as for the 
Higgs doublets $H_u$ and $H_d$ as discussed in~\cite{Peinado:2019mrn}. Moreover, to serve our purpose, we also include SM singlets $ N_R, N_L, \zeta$ with different PQ charges and an extra SM doublet $  \eta_{u} $ (see Table \ref{tab:scoto} for details).

Authors of Ref.~\cite{Chen:2012baa} have studied the Dirac neutrino masses generated at one-loop level within the PQ symmetric model by using charged leptons and charged extra scalars.  %
It is to be noted that Ref.~\cite{Carvajal:2018ohk} has  discussed the generation of the Dirac neutrino mass at one-loop level within the framework of scotogenic model in the context of KSVZ \cite{Kim:1979if,Shifman:1979if}. In both the cases the Diracness of neutrinos is guaranteed by Lepton number itself and not by the PQ symmetry as pointed out here. 
In our study, in both frameworks,  we show how it leads to a novel class of minimal axion models that effectively imply Dirac neutrinos.

\section{Framework}\label{sec:Frame}
It has been pointed out in~\cite{Peinado:2019mrn} that one can explain Dirac nature of neutrinos in the context of the DFSZ axion  model.
If we consider the model in Table ~\ref{originalDFSZ}, there is no way to generate a $\Delta L=2$ operator at any order at the perturbative level. 
If we include the RH neutrino transforming as a $-1$ as the rest of the RH fields, a Dirac mass is generated through the Yukawa coupling with the Higgs, in such a case, the Yukawa Lagrangian is given by 

\begin{eqnarray}\label{Lag1}
 {\cal L}_{Y} & = & y_{ij}^{u}\bar{Q}_{i}H_{u}u_{j} + y_{ij}^{u} \bar{Q}_{i}H_{d} d_{j} + y_{ij}^{l}\bar{L}_{i}H_{d}l_{j}  
 +     y_{ij}^{\nu}\bar{L}_{i}H_{u} \nu_{Rj} + h.c.~\;,
\end{eqnarray}
in such a way that the Yukawa couplings $y_{{ij}}^{\nu}$ must be ${\cal O}(10^{-12})$ in order to account for the recent KATRIN bound~\citep{Aker:2019uuj}.
 As was shown in ~\cite{Peinado:2019mrn}, if the RH neutrino transforms as $-5$ the direct Yukawa coupling is forbidden \footnote{A charge of +3 is also possible but in that case a Dirac mass is generated by the $\tilde{H_d}$ field.}. The effective  dimension-5 operator can be generated  as follows
 
\begin{equation}\label{Dim5Op}
 {\cal L}_{{dim~5}}^{D} =  y_{ij}^{\nu}\bar{L}_{i}H_{u} \nu_{Rj}\frac{\sigma}{\Lambda_{UV}}  + h.c.  \;, 
\end{equation}

where $ \sigma $ can be expressed as
\begin{equation}
\sigma(x) = \dfrac{1}{\sqrt{2}} \left( \rho(x) + f_a \right)  e^{i a(x) /  f_a }\;.
\end{equation}
Here,  $ a(x) $ is  the QCD axion \cite{Wilczek:1977pj,Weinberg:1977ma},  $ f_a $ is the PQ breaking scale and $ \rho(x)  $ is the radial part that will gain a mass of order of the PQ symmetry breaking scale.

As was pointed out previously, it can be easily UV completed through a type-I Dirac seesaw. In the following sections we will give two other frameworks to UV complete such an operator.
\begin{widetext}
 \begin{center}
\begin{table}[t] 
\begin{tabular}{|c | c c c   c c| c c c|}
  \hline  
\hspace{0.4cm} Fields/Symmetry  \hspace{0.4cm} & \hspace{0.4cm} $Q_i $ \hspace{0.4cm}  & \hspace{0.4cm} $ u_i$ \hspace{0.4cm}  & \hspace{0.4cm} $d_i$ \hspace{0.4cm}   & \hspace{0.4cm}  $L_i$  \hspace{0.4cm}  & \hspace{0.4cm} $l_i$ \hspace{0.4cm} &  \hspace{0.4cm} $H_u$ \hspace{0.4cm} & \hspace{0.4cm} $H_d$ \hspace{0.4cm} & \hspace{0.4cm} $\sigma$ \hspace{0.4cm} \\
\hline
$SU(2)_L\times U(1)_Y$   &  (2,1/6)  &  (1,2/3)  &  (1,-1/3)  &  (2,-1/2)  &  (1,-1)    &        (2,-1/2)                &      (2,1/2)               &          (0,0)           \\   
\hline
$U(1)_{PQ}$    &  1     &   -1     &  -1     &  1     &  -1    &         2               &     2              &         4             \\
\hline
  \end{tabular}
\caption{Quantum numbers in the DFSZ axion model.}\label{originalDFSZ}
\end{table}
\end{center}
\end{widetext}


\subsection{Case-I: Type-II Dirac seesaw}
Here we will consider a concrete example of an extension of the model in Table~\ref{originalDFSZ} 
by including the RH neutrino transforming as $-5$ under the PQ symmetry and an extra $SU(2)_L$ doublet, $\Phi_u$, with PQ
charge 6, see Table ~\ref{TII}. 
In this case there are terms in the potential of the form
\begin{equation}
V \sim \kappa H_{u} H_{d} \sigma^* + \lambda^{\prime}H_{d} \Phi_u \sigma^{*2} \;,
\label{eq:scalarpot}
\end{equation}
where the couplings $\kappa$ and $\lambda^{\prime}$ are  dimensionful and dimensionless, respectively.
Now, by inspecting the PQ charge assignments of different fields content as given in Table ~\ref{TII}, we notice that there is no way to form the dimension-5 Weinberg operator for the light neutrino masses, nor any other operator with powers of $\sigma$ and or powers of $H_{u}$, $H_{d}$ and $ \Phi_u $. 
We first write down the possible dimension-5 Weinberg operators in the presence of Higgs doublets
\begin{equation}\begin{array}{lcr}
&{\mbox{Operator}}&\mbox{PQ charge}\\\\
{\mathcal L}_{dim~5}\sim& 
\left\{\begin{array}{l}
\frac{LL\tilde{H}_{u} \tilde{H}_{u} }{\Lambda_{UV}}\\\\
\frac{LL\tilde{H}_{u} H_d }{\Lambda_{UV}}\\\\
\frac{LLH_d H_d }{\Lambda_{UV}} \\\\
\frac{LL\tilde{\Phi}_{u}\tilde{\Phi}_{u}}{\Lambda_{UV}} \\\\
\frac{LL\tilde{H}_{u} \tilde{\Phi}_{u} }{\Lambda_{UV}} \\\\
\frac{LLH_d \tilde{\Phi}_{u} }{\Lambda_{UV}} 
\end{array}\right.&\begin{array}{r}1+1+(-4)={-2}\\\\
1+1+(0)={+2}\\\\1+1+(4) = +6 \\\\1+1+(-12) = -10 \\\\1+1+(-8) = -6 \\\\1+1+(-4) = -2 \;.
\end{array}\end{array} \label{dim5}
\end{equation}
It is apparent from Eq. \eqref{dim5} that none of these operators are invariant under the PQ symmetry. Moreover, notice that all these operators transform as $m~mod(4)=2$ under PQ. Hence, from Eq. \eqref{dim5}, there is no way to construct an operator invariant under $\mathrm{U(1)_{PQ}}$ and the SM symmetries simultaneously.
We further realize that this argument also extends to all the higher order effective operators that could potentially generate Majorana neutrino masses.
In the following, we give all possible gauge invariant contractions of the scalar fields and their PQ charges:%
\begin{equation}\begin{array}{lrlr}
\sigma^n  & (4 n); & \quad
(\sigma^*)^n & (-4 n);\\
(H_u H_d)^n & (4 n);  & \quad
(H_u H_d)^{*n} & (-4 n);\\
(H_u^\dagger H_u)^n & (0); & \quad
(H_d^\dagger H_d)^n & (0);\\
(H_u^\dagger \Phi_u)^n & (4n); & \quad
(H_u^\dagger \Phi_u)^{*n} & (-4n); \\
(H_d \Phi_u)^n & (8n); & \quad
(H_d \Phi_u)^{*n} & (-8n); \\
(\Phi_u^\dagger \Phi_u)^n & (0);
\\\end{array}
\label{eq:pq-charge}
\end{equation}
%
where as we can see, all these contractions (and their combinations) are 0 or multiples of 4 under PQ symmetry. Therefore, there is no way to make a combination of operators on Eqs. \eqref{dim5} and \eqref{eq:pq-charge} invariant under PQ symmetry and hence, neutrinos must be Dirac particles.

The relevant part of Yukawa Lagrangian that generates Dirac neutrino masses is given by
\begin{equation}
\mathcal{L}_Y \supset y^{\nu}_{ij}\overline{\nu}_{Li} \Phi_u \nu_{Rj} + \mu H_{u}\Phi_u^{\dagger}\sigma + h.c.
\label{typeIlag}
\end{equation}

\begin{figure}
\centering
\includegraphics[scale=1]{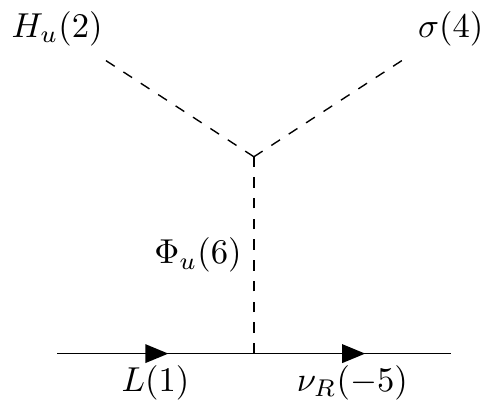}
\caption{\footnotesize Feynman diagram for Dirac neutrino masses in Type-II DFSZ scenario.}
\label{fig:type2diag}
\end{figure}

 \begin{table}[htb]
        \centering \scriptsize
             \begin{tabular}{|c|c|c|c|c|c|c| }
       \hline  
Symmetry/Fields &  $L_i$ &  $\nu_{Ri}$ & $H_{u}$  & $H_{d}$  & $\sigma$ & $\Phi_u$\\ 
       \hline
$SU(2)_L\times U(1)_Y$ & (2, -1/2)  & (1, 1)  & (2, -1/2) & (2, 1/2) & (0, 0)  & (2, -1/2)\\
       \hline
$U(1)_{PQ}$ & 1  &  -5  & 2 & 2 & 4 & 6 \\%
  \hline    
     \end{tabular}
     \caption{\footnotesize Fields content and transformation properties under PQ symmetry in type-II seesaw framework.}         \label{TII}

      \end{table} 


The scalar potential of the model contains the term $ \mu_\Phi^2 \Phi_u^\dagger \Phi_u  $, which in the small scalar mixing limit sets the mass scale for a heavy scalar $\Phi'$, composed mostly of $\Phi_u$. That is to say, the scalar particles of the model mix through a unitary matrix $K$ into a mass eigenstate basis as
\begin{equation}
\phi_i = K_{ij} S_j, 
\end{equation}
where $\phi$ are the real neutral components of the scalars $H_u$, $H_d$, $\Phi_u$ and $\sigma$. We consider the limit where one of the mass eigenstates is mostly composed of $\Phi_u$, with a mass $M_{\Phi_u}^2 >> v_u^2$. The largest contribution to the mixing between $\Phi_u$ and other fields is the $\mu$ term of Eq. \eqref{typeIlag}, the large vev of $\sigma$ can induce a large mixing between $H_u$ and $\Phi_u$. This mixing can raise the mass of one of the light eigenstates above the EW scale, excluding the possibility that the scalar which is predominantly $H_u$ is the 125 GeV Higgs boson. At leading order the mixing between these fields goes as
\begin{equation}\label{eq:scalarmixing}
 \sin\theta \sim \dfrac{\mu f_a}{M_{\Phi_u}^2} \;.
\end{equation}
Therefore, the smallness of $\theta$ demands $\mu f_a<<M^2_{\Phi_u}$. This condition is similar to the fine-tuning of $\kappa$ in Eq. (\ref{eq:scalarpot}), needed to separate the PQ scale from the EW scale, as mentioned in \cite{Langacker:1986rj}. Furthermore, the fine-tuning of these parameters has been shown to be stable under radiative corrections once stablized at the tree-level \cite{Clarke:2015bea}.
The neutrino masses resulting from the breaking of the $SU(2)_L\times U(1)_Y$ and $U(1)_{PQ}$ symmetries by the scalar vevs  $\langle H_u\rangle=v_{u}$ and $\langle  \sigma\rangle=f_a$ are
\begin{equation}\label{eq:mnuCase1}
(m_\nu)_{ij}=y^\nu_{ij} \frac{\mu v_u f_a }{M_{\Phi_u}^2} \sim y^\nu_{ij} v_u \sin\theta  \;,
\end{equation}
where in the last term we have used Eq. \eqref{eq:scalarmixing}.

\begin{figure}
\includegraphics[height=6cm, width=8cm]{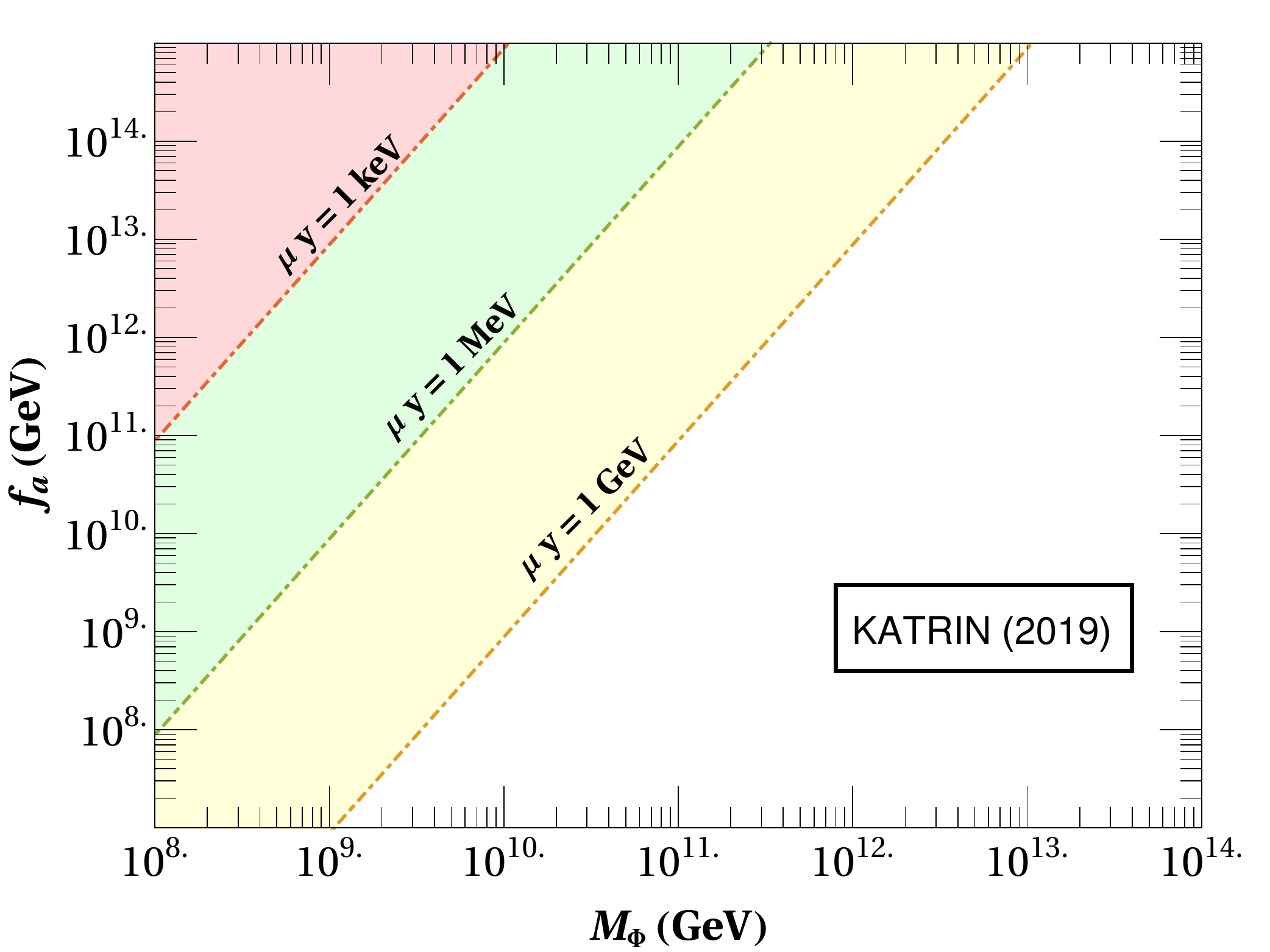}
\includegraphics[height=6cm, width=8cm]{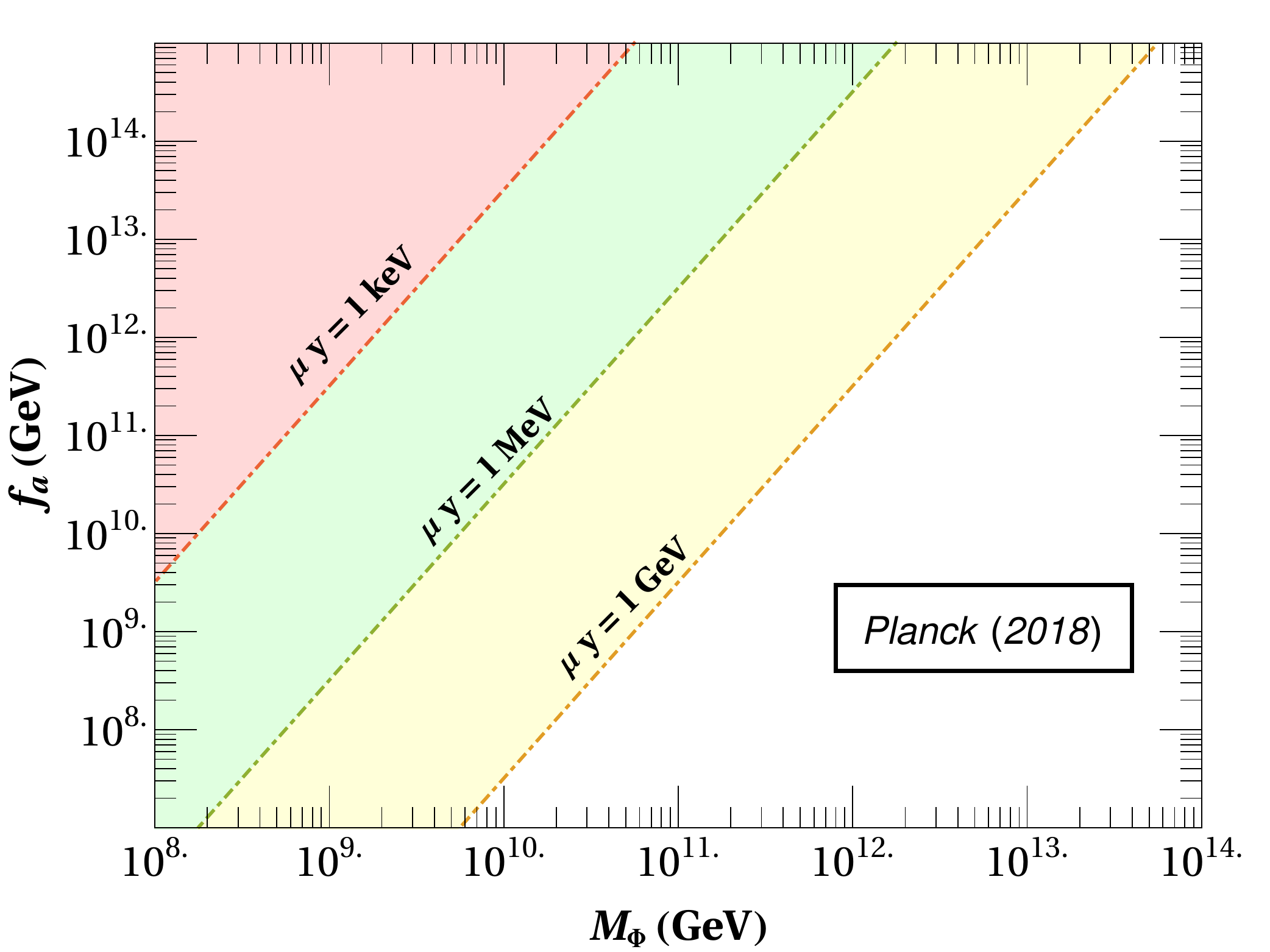}
\caption{\footnotesize Exclusion region plots (colored regions are excluded) in ($M_{\Phi_u} -  f_a$) plane for a type-II Dirac seesaw mechanism. Three benchmark values for $\mu y = 1 {\rm GeV}, 1 {\rm MeV}, 1 {\rm keV}$ have been adopted, respectively. The plots are presented by using the limits on neutrino mass from KATRIN~\citep{Aker:2019uuj} which gives $ m_\nu <  1.1$ eV at 90\% C.L.  (left panel) and Planck  \cite{Aghanim:2018eyx} $\sum m_\nu <0.12$ eV (at 95 \% confidence level using TT, TE, EE + lowE + lensing + BAO) (right panel). }
\label{fig:typeIspace}
\end{figure}

In the type-I Dirac seesaw scenario, as pointed out in \cite{Peinado:2019mrn}, a large hierarchy among the PQ scale and the mediator scale is needed in order to explain the tiny neutrino masses. However, here  the dependency  is on the inverse squared mass. This suggests that a smaller mass hierarchy than in the type-I seesaw~\cite{Peinado:2019mrn} may be allowed. As can be seen from Eq. (\ref{eq:mnuCase1}), the smallness of neutrino mass is required by the smallness of the scalar mixing angle $\theta$.
The measurement of the tritium beta decay spectrum at KATRIN~\citep{Aker:2019uuj} currently yields a direct limit for neutrino masses of $ m_\nu <  1.1$ eV at 90\% C.L. , while the indirect limit from Cosmological measurements \cite{Giusarma:2016phn,Vagnozzi:2017ovm,Aghanim:2018eyx} constrains them further to $\sum m_\nu <0.12$ eV (at 95 \% confidence level using TT, TE, EE + lowE + lensing + BAO). The bounds of tritium beta decay and cosmology are translated into bands for the allowed scales for $f_a$ and $M_\phi$ as shown in Fig.~\ref{fig:typeIspace}.

\subsection{Case-II: One-loop Dirac seesaw}
In this section, we discuss a one-loop mechanism to UV complete the effective  coupling of Eq.~\eqref{Dim5Op}.
A detailed discussion of all the possible topologies to explain Dirac  neutrino  masses with four external lines was outlined  in~\cite{Bonilla:2019hfb}. In what follows, we consider the most economical scenarios for the dimension-5 operator which can lead to Dirac  neutrino  masses.
All the necessary fields carry $SM\otimes PQ$  charges are presented in Table (\ref{tab:altloop}). Under this assignment of PQ charges, and the subsequent spontaneous symmetry breaking, the residual symmetry is $Z_2$. 
Note that the leptons, quarks, $\eta_u$ and $\zeta$ are odd under the $Z_2$ residual symmetry. Stability is achieved for the lightest odd scalar or for $N$ by the interplay of $Z_2$ and Lorentz invariance \cite{Bonilla:2018ynb} \footnote{We also provide an alternate loop model considering half-integral PQ charges for the particles running inside the loop in appendix~\ref{sec:app}. There dark matter stability is obtained under $Z_4$  residual symmetry.}. This stability results in two scenarios with a multicomponent Dark Matter, a fraction $\Omega_a$ composed by the axion, and another fraction $\Omega_{\rm WIMP}$ composed by the stable WIMP, such that \cite{Aghanim:2018eyx}
\begin{equation}
\Omega_{\rm CDM} h^2 (= 0.12) \geq (\Omega_{a}+\Omega_{\rm WIMP})h^2.
\end{equation}
While the relic density of WIMP dark matter is determined by the thermal freeze-out mechanism and can be calculated from the parameters of the model, the density of axions is determined from other production mechanisms such as the axion misalignment mechanism or from topological defects of the axion field \cite{Duffy:2009ig,DiLuzio:2020wdo}. The resulting relic density from these mechanisms is highly dependent on the cosmological history of the axion field, and on the precise cosmological scenario considered. An analysis of these mechanisms is beyond the scope of this work.
\begin{table}[htb]
        \centering \scriptsize
       \begin{tabular}{|c|c|c|c|c|c|c|c|c|c| }
       \hline  
Symmetry/Fields &  $L_i$ &  $\nu_{Ri}$ & $H_{u}$  & $H_{d}$ & $ N_R $ & $ N_L $ & $\sigma$ & $\eta_u$ & $ \zeta $\\ 
       \hline
$SU(2)_L\times U(1)_Y$ & (2, -1/2)  & (1, 1)  & (2, -1/2) & (2, 1/2) & (1, 0) & (1, 0) & (1, 0)  & (2, -1/2) & (1, 0)\\
       \hline
$U(1)_{PQ}$ & 1  &  -5  & 2 & 2 & 2 & 2  & 4 & -1 &  7\\
		\hline
$Z_2^{PQ}$ & -1 &  -1  & +1 & +1 & +1  & +1  & +1 & -1 &  -1 \\%
  \hline    
     \end{tabular}
     \caption{\footnotesize Field content and transformation properties under PQ symmetry in the alternative one-loop mechanism.}  
     \label{tab:altloop}     
\end{table} 
\begin{figure}
\includegraphics[scale=1]{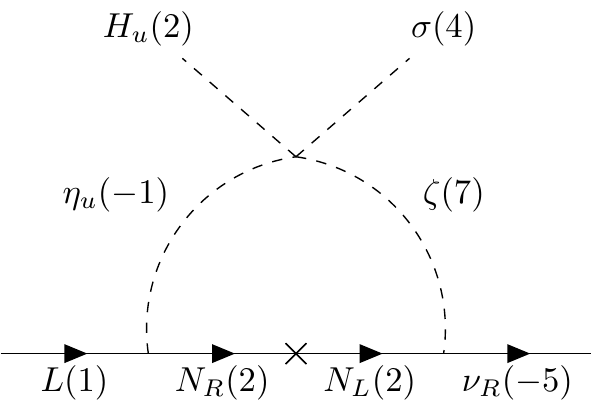}
\includegraphics[scale=0.25]{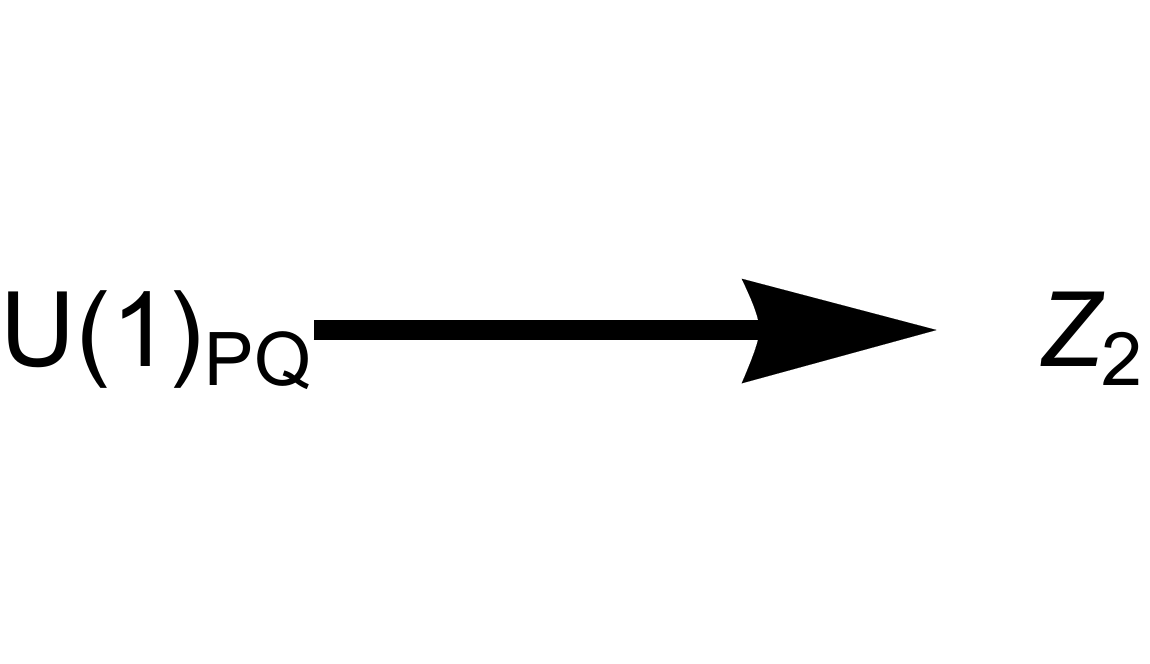}
\includegraphics[scale=1]{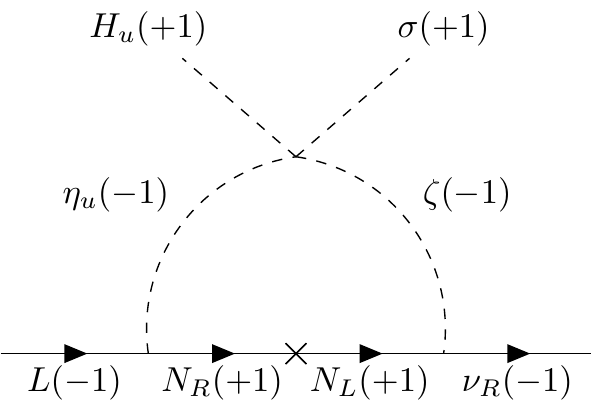}
\caption{\footnotesize Feynman diagram for Dirac neutrino masses in alternative one-loop DFSZ scenario. Here left panel respects PQ charge assignment, whereas right panel respects the remnant $Z_2^{PQ}$ charge assignment that  arises due to PQ symmetry breaking.}
\label{fig:loopaltdiag}
\end{figure}

Now we write down the scalar potential that is allowed by the $SM\times U(1)_{PQ}$ symmetries as below,
\begin{align}
V  = ~ & \mu^{2}_{u} H_{u}^\dagger H_{u} + \lambda_u (H_{u}^\dagger H_{u})^2 
 + \mu^{2}_{\eta_u} \eta_{u}^\dagger \eta_{u} + \lambda_{\eta_u} (\eta_{u}^\dagger \eta_{u})^2  \\ \nonumber
 +~  & \mu^{2}_{\sigma} \sigma^* \sigma  + \lambda_\sigma (\sigma^* \sigma)^2  + \mu^{2}_{\zeta} \zeta^* \zeta  + \lambda_\zeta (\zeta^* \zeta)^2   \\ \nonumber
 %
 %
+ ~ & \lambda^1_{u\eta_u} (H_{u}^\dagger H_{u}^{}) (\eta_{u}^\dagger \eta_{u})   + \lambda^2_{u\eta_u} (H_{u}^\dagger \eta_{u}^{}) (\eta_{u}^\dagger H_{u}) 
+    \lambda_{u \sigma} (H_{u}^\dagger H_{u})( \sigma^* \sigma ) \\ \nonumber
+ &   \lambda_{u \zeta} (H_{u}^\dagger H_{u})( \zeta^* \zeta)
+  \lambda_{\eta_u \zeta} ~ (\eta_{u}^\dagger \eta_{u}) (\zeta^* \zeta) +  \lambda_{\eta_u \sigma} ~ (\eta_{u}^\dagger \eta_{u}) (\sigma^* \sigma)  +  \lambda_{\sigma \zeta} ~ (\sigma^* \sigma )(\zeta^* \zeta) \\ \nonumber
 + & \kappa H_{u} H_{d} \sigma^* + \lambda_1 [H_{u} \eta^\dagger_u \zeta^{*} \sigma + h.c. ]  \;. 
\end{align}
Note that all the terms that are allowed for $ H_{u} $ are also allowed for $ H_{d} $ except for $\lambda^2_{u\eta_u}, \lambda_1 $ terms, which are not invariant under the SM gauge group.
After the electroweak symmetry breaking, the SM Higgs doublet $ H_{u} $ acquires its vev $v_u$. Then, considering the $\eta_u-\zeta$ mixing matrix, the mass matrix for the neutral scalars in  $(\eta^0_u, \zeta)$ basis can be written as
\begin{equation}\label{eq:DMMass}
\mathcal{M} = 
\begin{pmatrix}
\mu^{2}_{\eta_u} +   \lambda_{u\eta_u} v^2_u + \lambda_{\eta_u \sigma} f^2_a & ~~   \lambda_1 v_u f_a\\
\lambda_1 v_u f_a & ~~ \mu^{2}_{\zeta} +  \lambda_{u \zeta}  v^2_u + \lambda_{\sigma \zeta} f^2_a
\end{pmatrix} \;.
\end{equation} 
where $f_a$ is the PQ symmetry breaking scale and  we used $\lambda_{u\eta_u}  = \lambda^1_{u\eta_u} + \lambda^2_{u\eta_u}$.
We describe the relevant part of Yukawa Lagrangian for the leptons as
\begin{equation}
\mathcal{L}_Y \supset y^{\nu}_{ij}\overline{L}_{i} \eta_u N_{Rj} + M_{jk} \overline{N}_{Rj} N_{Lk}+ y^{\nu \prime}_{ki}\overline{N}_{Lk} \nu_{Ri} \zeta + h.c.
\end{equation}

The neutrino mass obtained from the one-loop diagram in Fig. \ref{fig:loopdiag} is given by \cite{Ma:2006km,Carvajal:2018ohk}
\begin{equation}
m_\nu^{ij}= \frac{1}{64 \pi^2} \sum_{X=R,I}\frac{\lambda_1 v_u f_a}{m_{S_{X_2}}^2- m_{S_{X_1}}^2} \sum_{k}y^{ik}y'^{kj} m_{N_{k}} \left[  F\left(\frac{m^2_{S_{X_2}}}{m^2_{N_i}} \right) -F \left(\frac{m^2_{S_{X_1}}}{m^2_{N_i}} \right) \right] \;,
\end{equation}
where $F(x)=x\log (x)/(x-1)$ and $S_{R_i/I_i}$ (for $i = 1, 2$) denote the CP even and odd mass scalar eigenstates, obtained from the $\eta_u-\zeta$ mixing.
For the fermionic WIMP case, considering $m^2_{S_1}\sim m^2_{S_2} =m_S^2 >> \lambda_1 f_a v_u $ , and subsequently in the $m_{N_{light}} << m_S$ limit, the neutrino mass matrix can be expressed as
\begin{eqnarray} \label{eq:Mass1loop}
m_\nu^{ij} & \sim & \frac{\lambda_1 v_uf_a}{32 \pi^2} \sum_{k} y^{ik} y'^{kj}  
\frac{m_{N_k}}{m_{S}^2} \;. 
%
\end{eqnarray}
To find dark sector scalar masses i.e., the mass eigenvalues of Eq.~\eqref{eq:DMMass},  we make a simple numerical estimation. 
We take $ v_u \sim 100 $ GeV, $ f_a \sim 10^{12} $ GeV and $ \lambda_1 = 10^{-7} $ in our calculation.
We write down $ \mathcal{M} $ as
\begin{equation}\label{eq:NumDMMass}
\mathcal{M} = 
\begin{pmatrix}
10^{24} +   10^{-5} \times 10^{4} - 10^{- 4} \times 10^{24} & ~~   10^{-7} \times 10^{2}\times 10^{12} \\
* & ~~ 10^{20} +  0.92 \times 10^{4} - 10^{- 4} \times 10^{24}
\end{pmatrix} [ \rm GeV^{2}]\;.
\end{equation} 
For simplicity,  we assume $  \lambda_{\eta_u \sigma} =  \lambda_{\sigma \zeta}$.
We also consider $ \mu_{\eta_u} = 10^{12}$ GeV, $ \mu_{\zeta} = 10^{10}$ GeV together with  $ \lambda_{u\eta_u} = 10^{-5} $, $\lambda_{u \zeta} = 0.92$ and $ \lambda_{\eta_u \sigma} = - 10^{-4} $.
Diagonalization of Eq.~\eqref{eq:NumDMMass}  leads to the lightest DM of $ \mathcal{O} (130) $ GeV, whereas the heaviest DM came out as $ \sim  10^{12}$ GeV for the given set of numerical values.
Now, considering these numerical values and using them in Eq.~(\ref{eq:Mass1loop}), one estimates $ \mathcal{O} (1) $ eV active neutrino masses as follows:
\begin{equation} \label{eq:Mass1loopN}
%
m_\nu^{ij}   \sim  1 {\rm eV} \left(\frac{\lambda_1}{10^{-7}}\right) \left(\frac{v_u}{10^2 \text{GeV}}\right)\left(\frac{f_a}{10^{12}\text{GeV}}\right)   \left(\frac{y}{10^{-7}}\right) \left(\frac{y'}{0.82}\right) \left(\frac{m_{N_k}}{130\text{GeV}}\right) \left(\frac{(280\text{GeV})^2}{m_{S}^2}\right)  \;.
\end{equation}
As we are considering a small left-handed lepton-dark sector coupling to avoid saturating the limits on Lepton Flavor violating processes, the dark matter relic density can be determined by annihilation of $N$ into right handed neutrinos. This process is absent from the Majorana neutrino Scotogenic model, and its thermally averaged cross section $\langle \sigma v \rangle$ is given by \cite{Ma:2019yfo}
\begin{equation}\label{eq:FermionicDM}
\sigma \times v_{rel} = \frac{y^{\prime 4}}{32 \pi^2} ~ \frac{m_{N_k}^2}{(m_{S}^2 +  m_{N_k}^2)^2} \;.
\end{equation}
Notice here that we are considering the annihilation channel $NN \rightarrow \nu \nu$ mediated by the dark sector singlet scalar, in the small scalar mixing limit.  To have a simple numerical estimation, we take $y^{\prime} = 0.82, m_{S}  = 280 ~{\rm GeV}$, and $m_{N_k} = 130$ GeV, which then leads to  $\sigma \times v_{rel}  \simeq 1$ pb and hence one finds the observed $\Omega h^2 \simeq$  0.12~\cite{Aghanim:2018eyx}. Thus in this case the axion plays no role in the dark matter relic density. Alternatively, by setting the Yukawa coupling $y'\sim 1$ lowers the relic density contribution of $N$ to a quarter of the total dark matter relic density, $\Omega_{N} \sim 0.4 \quad \Omega_{CDM}$. The remaining fraction of dark matter density may come from the axion relic density. Therefore we find that this model can accommodate axions as a negligible, or dominant form of dark matter.
Alternatively, considering the scalar WIMP Dark Matter case $m_S<<m_{N_k}$, we find
\begin{equation}
m_\nu^{ij} \sim \frac{\lambda_1 v_uf_a}{32 \pi^2} \sum_{k} \frac{y^{ik} y'^{kj}}  {m_{N_K}} \left[ \log \left(\frac{m_{N_k}^2}{m_S^2}\right)   -1  \right] \;.
\label{eq:MassScalarDM}
\end{equation}
We  find $ \mathcal{O} (1) $ eV order masses for neutrinos by setting $\lambda_1 \sim 10^{-5}$, $v_u \sim 10^2$ GeV, $f_a\sim 10^{12}$ GeV, $y \sim 10^{-5}$, $y' \sim 10^{-4}$, $m_N \sim 10^8$ GeV and $m_S \sim 1$ TeV.
Here, the DM is a mixture of a electroweak doublet and a singlet, as in \cite{Kakizaki:2016dza}. Two limiting cases are found in the small mixing regime, when DM is mostly composed of $\eta_u$ or of $\zeta$.
The mixings between $\eta_u$ and $\zeta$ are given, at leading order by 
\begin{equation}
\sin 2 \theta_X \sim \frac{\lambda_1 v_uf_a}{m_{S_{X_2}}^2- m_{S_{X_1}}^2}\;.
\end{equation}\\
We can obtain a $\sim$ TeV scale eigenstate from this mixing using the following assignment of parameters in Eq. (\ref{eq:DMMass}): $ \lambda_{u\eta_u} =  \lambda_{u \zeta}$,  $  \lambda_{\zeta \sigma} = 10^{-14} \lambda_{\eta_u \sigma}$, $ \mu_{\eta_u} = 10^{12}$ GeV, $ \mu_{\zeta} = 10^{6}$ GeV together with  $ \lambda_{u\eta_u} = 1 $ and $ \lambda_{\eta_u \sigma} = - 10^{-4} $.
When the lightest dark scalar is mostly a gauge singlet, $S_{light}\approx\zeta$, the annihilation channel for thermal production is through the scalar couplings in the potential \cite{McDonald:1993ex,Cline:2013gha}. 
This means that the annihilation processes $S_{light} S_{light} \rightarrow {\rm SM~SM}$ that determine the relic abundance of $S_{light}$ are mediated by scalar channels.\\
For example, the $S_{light} S_{light} \rightarrow h h$ annihilation channel contribution to $\langle\sigma v \rangle$ is given by \cite{McDonald:1993ex}
\begin{equation}
\langle\sigma v \rangle_{hh} = \frac{\lambda_{u \zeta}^2}{64 \pi m_{S}^2} \left(1- \frac{m_h^2}{m_S^2}  \right)^{1/2},
\end{equation}
where $h$ is the 125 GeV SM Higgs. The additional annihilation channels into the SM fermions and gauge bosons are also controlled by $\lambda_{u \zeta}$.
A scalar coupling $\lambda_{u \zeta}$ of order $\mathcal{O}(10^{-1})$ is required to account for $\Omega_{S} h^2 \sim 0.12$ at a scalar WIMP mass $m_S$ of 1 TeV \cite{Athron:2017kgt}. A scalar coupling $\lambda_{u\zeta} \sim 0.2$ \footnote{While a $\lambda_{u\zeta} \sim 0.2$ coupling would saturate the direct detection limit from XENON1T \cite{Aprile:2018dbl} at $\Omega_{S_{light}}=\Omega_{CDM}$, the diminished contribution of $S$ to the dark matter density relaxes this bound.} yields a relic density of $S_{light}$ approximately forty percent of the total dark matter density, leaving the axion as the dominant component.  \\
When the lightest dark scalar is mostly a doublet, $S_{light}\approx\eta_u$, additional channels mediated by gauge couplings appear. Masses of $\mathcal{O} (1)$ TeV are compatible with the relic density and direct detection constraints, using the quartic scalar couplings with the SM Higgs of order $\mathcal{O}(0.1)$ \cite{Cirelli:2005uq, Kalinowski:2018kdn}.  
Given the possibility of having mixed dark matter in the model, a major difference of this model with ``pure" models is that the lower relic density of the WIMP DM candidate needed to acommodate the axion requires larger couplings to the SM. This results in a larger direct detection signals for their relatively smaller densities.
For example, the DM direct detection experiments \cite{Akerib:2017kat} constrain only the WIMP component of DM, while the axion detection experiments \cite{Braine:2019fqb} constrain the axion component. On the other hand, the decreased abundance of the WIMP component of DM necessitates larger couplings for it to augment the annihilation cross section. This same couplings are involved in Lepton Flavor Violating (LFV) processes, such as the $\mu \rightarrow e \gamma$ decay, which are strongly constrained. In this model, the additional interaction with $\nu_R$ may be exploited to enhance the annihilation rate while keeping the LFV inducing couplings low. For example, the $\mu \rightarrow e \gamma$ branching ratio is given by \cite{Toma:2013zsa}
\begin{equation}
\textbf{Br}(\mu \rightarrow e \gamma) = \frac{3 \alpha_{em} \textbf{Br}(\mu \rightarrow e \bar{\nu_e} \nu_\mu)}{64 \pi G_F^2 m_{\eta^+}^4} \left|  \sum_k y_{e k}y^*_{\mu k} G\left( \frac{M_{N_k}^2}{m_{\eta^+}^2} \right)  \right|^{2},
\end{equation}
where $G_F$ is the Fermi constant, $\alpha_{em} $ is the fine structure constant and $\eta^+$ is the charged scalar from the $\eta_u$ doublet. The loop function $G(x)$ is defined by 
\begin{equation}
G(x)= \frac{1-6x+3x^2+2x^3-6x^2\ln x}{6(1-x)^4} \;.
\end{equation}
In the fermionic WIMP case, using the parameters we have provided in Eq. (\ref{eq:Mass1loopN}) and assuming $m_\eta^+\sim m_{S_{1}}$ we obtain a branching ratio of the $\mu \rightarrow e \gamma$ decay of order $\sim 10^{-33}$, well below the experimental bound $\textbf{Br}(\mu \rightarrow e \gamma) \leq 4.2 \times 10^{-13}$ \cite{TheMEG:2016wtm}. The new annihilation channel of the fermionic WIMP dark matter can result in an increased production of right handed neutrinos in the early Universe, which may oversaturate the effective number of relativistic degrees of freedom, $N_{eff}$ \cite{Zhang:2015wua}. This may disfavour the fermionic DM case.  

\section{Summary}\label{sec:summary}
We have discussed the DFSZ model where neutrinos are Dirac particles due to the PQ symmetry~\cite{Peinado:2019mrn}. In this context, we propose two different scenarios to generate naturally small effective Yukawa coupling for neutrinos. In order to explain the smallness of the Yukawa coupling, the tree-level coupling is forbidden by the PQ symmetry while an effective dimension-5 operator with the PQ field is allowed. This means that the Dirac neutrino mass is proportional to the PQ breaking scale.  
The first scenario is based on the Type-II Dirac seesaw where an extra heavy $SU(2)_L$ doublet allows Dirac neutrino mass when acquires a small vev once the PQ and the EW symmetry are broken. We summarize our results for this scenario in Fig.~\ref{fig:typeIspace}, considering the latest KATRIN~\citep{Aker:2019uuj} and the Planck data \cite{Aghanim:2018eyx}. These constraints set limits on the PQ breaking scale $f_a$ and the mass of the heavy scalar.

We also discuss the UV completion of the dimension-5 operator at one-loop level. In this context once the PQ is broken, a   residual $Z_2$ symmetry remains. The SM fermions are odd while the scalars are even, making the lightest field inside the loop stable by means of the residual $Z_2$ and Lorentz invariance giving a potential DM candidate. This residual symmetry is crucial, otherwise the the scalar particles inside the loop (odd under $Z_2$) acquire a vev,  and the loop would be a correction to the type-I Dirac seesaw. Therefore, in this scenario we have a potentially rich phenomenology with two dark matter components, a stable WIMP running inside the neutrino mass loop and the axion.  We have also discussed what are the parameters of such a dark sector in order to avoid an overclosed Universe.
It is worth mentioning that in both cases, the UV scale is more relaxed that in the type-I case~\cite{Peinado:2019mrn}, where the KATRIN neutrino bound set it to be ${\cal O}(M_{GUT})-\cal{O}(M_{\rm PLANCK})$.

\section{Acknowledgements}

This work is supported by the  grants DGAPA-PAPIIT IN107118 (M\'exico), CONACYT CB-2017-2018/A1-S-13051 (M\'exico) and  the  German-Mexican  research  collaboration grant SP 778/4-1 (DFG) and 278017 (CONACYT). NN is supported by the postdoctoral fellowship program DGAPA-UNAM. We thank Mario Reig for useful comments. LMGDLV is supported by CONACYT National Scholarships program.

\appendix
\section{An Alternate Loop Model}\label{sec:app}
%
In this appendix we present an alternate loop model to generate Dirac neutrino mass.
The $SM\otimes PQ$  charges of all the necessary fields are presented in Table \ref{tab:scoto}.


 \begin{table}[htb]
        \centering \scriptsize
       \begin{tabular}{|c|c|c|c|c|c|c|c|c|c| }
       \hline  
Symmetry/Fields &  $L_i$ &  $\nu_{Ri}$ & $H_{u}$  & $H_{d}$ & $ N_R $ & $ N_L $ & $\sigma$ & $\eta_u$ & $ \zeta $\\ 
       \hline
$SU(2)_L\times U(1)_Y$ & (2, -1/2)  & (1, 1)  & (2, -1/2) & (2, 1/2) & (1, 0) & (1, 0) & (1, 0)  & (2, -1/2) & (1, 0)\\
       \hline
$U(1)_{PQ}$ & 1  &  -5  & 2 & 2 & 1/2 & 1/2  & 4 & 1/2 &  11/2\\%
  \hline    
     \end{tabular}
     \caption{\footnotesize Fields content and transformation properties under PQ symmetry in the one-loop mechanism.}  
     \label{tab:scoto}     
      \end{table} 
%
%
\begin{figure}
\includegraphics[scale=1]{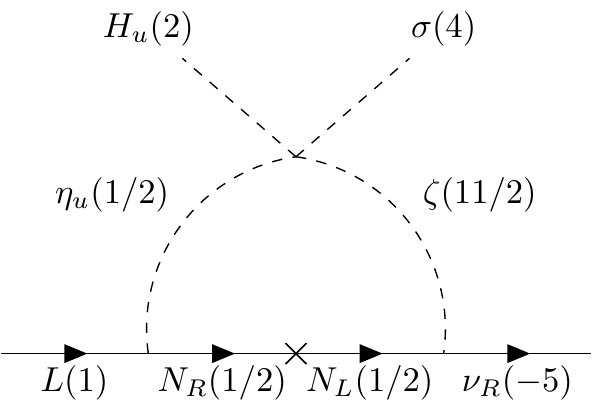}
\vspace{1cm}
\includegraphics[scale=0.25]{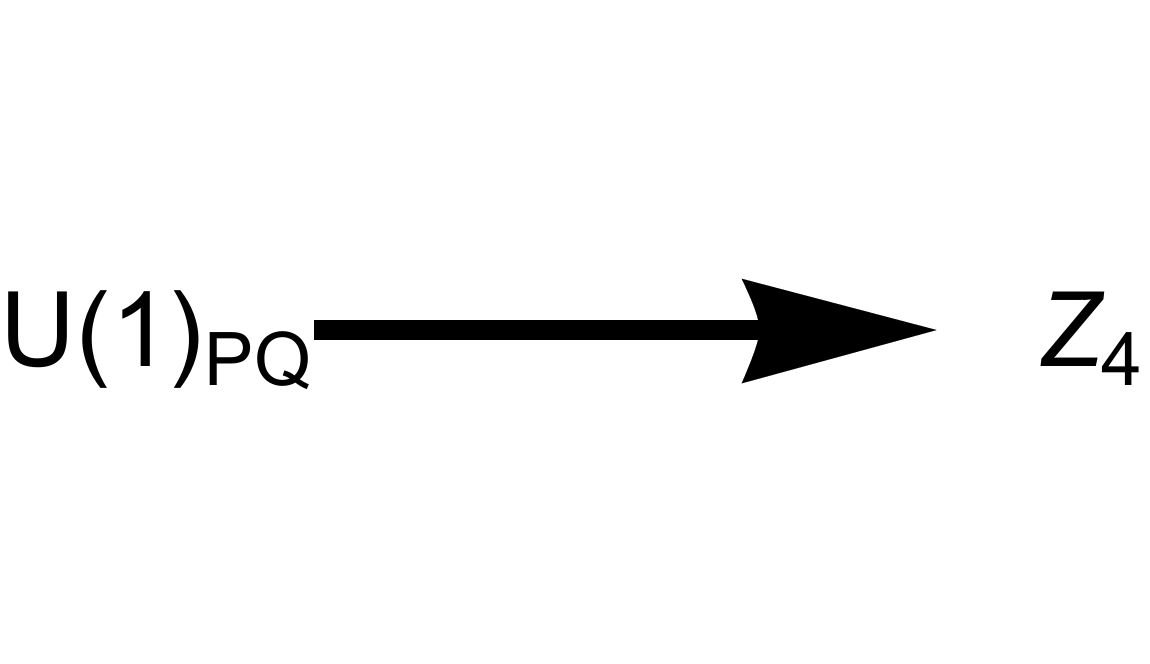}
\includegraphics[scale=1]{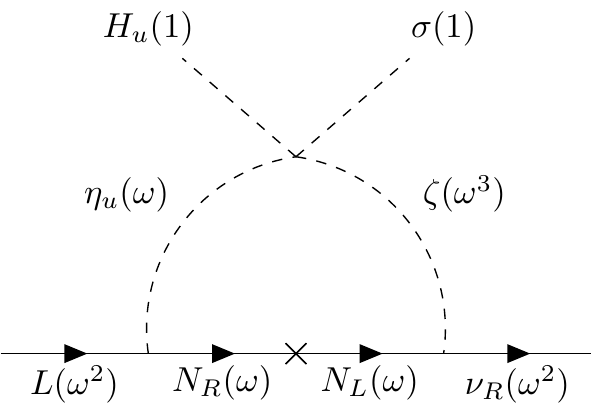}
\caption{\footnotesize Feynman diagram for Dirac neutrino masses in one-loop DFSZ scenario. Here left panel respects PQ charge assignment, whereas right panel respects remnant $Z_4^{PQ}$ charge assignment,  arises due to PQ symmetry breaking.}
\label{fig:loopdiag}
\end{figure}

In this case, the vevs of $\sigma$, $H_u$ and $H_d$ break the PQ symmetry into a $Z_4$ symmetry. The fields' transformation rules are given  in Table \ref{breaking}. Here, the particles inside the loop are automatically stable~\cite{Bonilla:2018ynb}.

 \begin{table}[htb]
        \centering \scriptsize
       \begin{tabular}{|c|c|c|c|c|c|c|c|c|c| }
       \hline  
Symmetry/Fields &  $L_i$ &  $\nu_{Ri}$ & $H_{u}$  & $H_{d}$ & $ N_R $ & $ N_L $ & $\sigma$ & $\eta_u$ & $ \zeta $\\ 
       \hline
$SU(2)_L\times U(1)_Y$ & (2, -1/2)  & (1, 1)  & (2, -1/2) & (2, 1/2) & (1, 0) & (1, 0) & (1, 0)  & (2, -1/2) & (1, 0)\\
       \hline
$Z_4^{PQ}$ & $\omega^2$ &  $\omega^2$  & 1 & 1 & $\omega$& $\omega$  & 1 & $\omega$ &  $\omega^3$ \\%
  \hline    
     \end{tabular}
     \caption{\footnotesize Fields content and transformation properties under PQ symmetry in the one-loop mechanism.}  
     \label{breaking}     
      \end{table} 

 We notice from the right panel of Fig. (\ref{fig:loopdiag}) that all the particles inside the loop carry $Z_4$ odd charges, whereas the SM particles are even under $Z_4$. Therefore, one can see that any combination of SM fields will be even under the $Z_4$
charges. Further forbidding all effective operators to dark matter decay.

\bibliography{bibliography}
\end{document}